\documentclass[11pt]{article} 
\usepackage{appb,epsfig} 
\newcommand{\bm}[1]{\mbox{\boldmath $#1$}}
\newcommand{\be}{\begin{equation}}
\newcommand{\ee}{\end{equation}}
\newcommand{\bea}{\begin{eqnarray}}
\newcommand{\eea}{\end{eqnarray}}
\newcommand{\st}{{\scriptscriptstyle T}}
\newcommand{\xbj}{x_{\scriptscriptstyle B}}
\newcommand{\zh}{z_h}
\def\slash{\rlap{/}}
\begin{document}

\title{
\begin{flushright}
\small
VUTH 98-09, hep-ph/9803230
\end{flushright}
\mbox{}\\
SINGLE SPIN ASYMMETRIES IN SEMI-INCLUSIVE DEEP INELASTIC 
SCATTERING\thanks{Presented at The Cracow Epiphany Conference on 
Spin Effects in Particle Physics, Cracow, Poland, January 9-11, 1998.}
}

\author{P.J. MULDERS
\address{Department of Physics and Astronomy, Vrije Universiteit,\\
De Boelelaan 1081, NL-1081 HV Amsterdam, the Netherlands}
}
\maketitle 
\begin{abstract}
In this talk I want to illustrate the many possibilities for studying
the structure of hadrons in hard scattering processes by giving a number
of examples involving increasing complexity in the demands for particle
polarization, particle identification or polarimetry. In particular
the single spin asymmetries will be discussed.
The measurements discussed in this talk are restricted to lepton-hadron
scattering~\cite{MT96,BM98}, but can be found in various other hard processes 
such as Drell-Yan scattering~\cite{TM95,BMT98} 
or $e^+e^-$-annihilation~\cite{BJM97}. 
\end{abstract}
  
\section{Introduction}

In inclusive deep inelastic lepton-hadron scattering (DIS) one is familiar 
with the factorization of the cross section, schematically
\begin{equation}
\sigma^{eH\rightarrow eX} = \sum_q f^{H\rightarrow q} \otimes
\sigma^{eq\rightarrow eq},
\end{equation}
which can be justified via the operator product expansion. Restricting
ourselves to quarks one finds local operators of the form $\overline \psi\,
D\ldots D\,\psi$ to be important, which can be resummed into nonlocal
operators $\overline \psi(0)\,\psi(x)$, in which the nonlocality is
restricted along the lightcone. In the case of inclusive scattering 
transverse momenta are irrelevant.
In semi-inclusive deep inelastic lepton-hadron scattering (SIDIS) 
the factorization
\begin{equation}
\sigma^{eH\rightarrow ehX} = \sum_q f^{H\rightarrow q} \otimes
\sigma^{eq\rightarrow eq} \otimes D^{q\rightarrow h},
\end{equation}
is much less well founded. There is no operator product expansion for
the process, but one starts with a hard scattering approach.
Furthermore, transverse momenta do matter in this process. The
soft parts, the distribution function $f$ and the fragmentation function
$D$ involve not only operators $\overline \psi\, D\ldots D\,\psi$,
but also operators of the form $\partial(\overline \psi\, D\ldots D\,\psi$),
which implies, when organized into nonlocal operators 
$\overline \psi(0)\,\psi(x)$, that the transverse separation becomes
important, although the separation remains lightlike.

\begin{figure}
\centerline{
\epsfig{figure=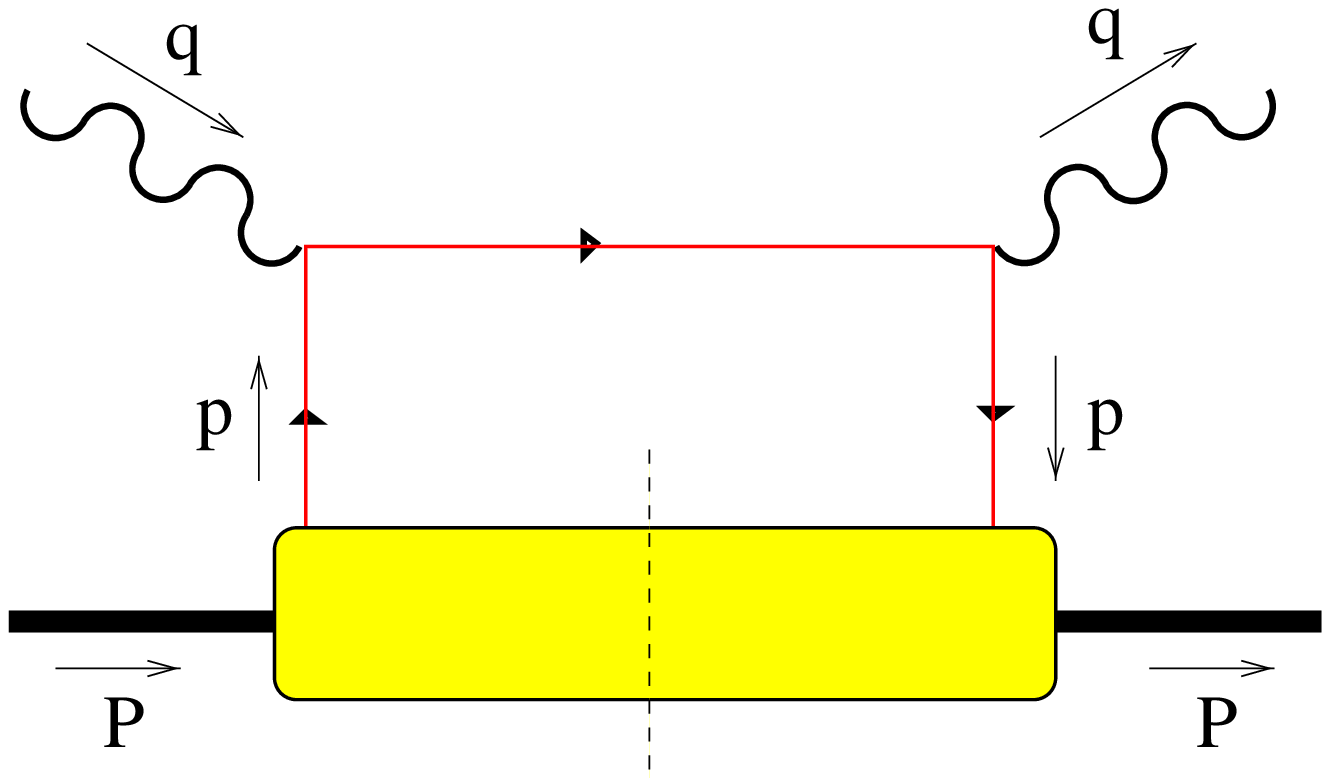,width=5cm}
\hspace{2cm}
\epsfig{figure=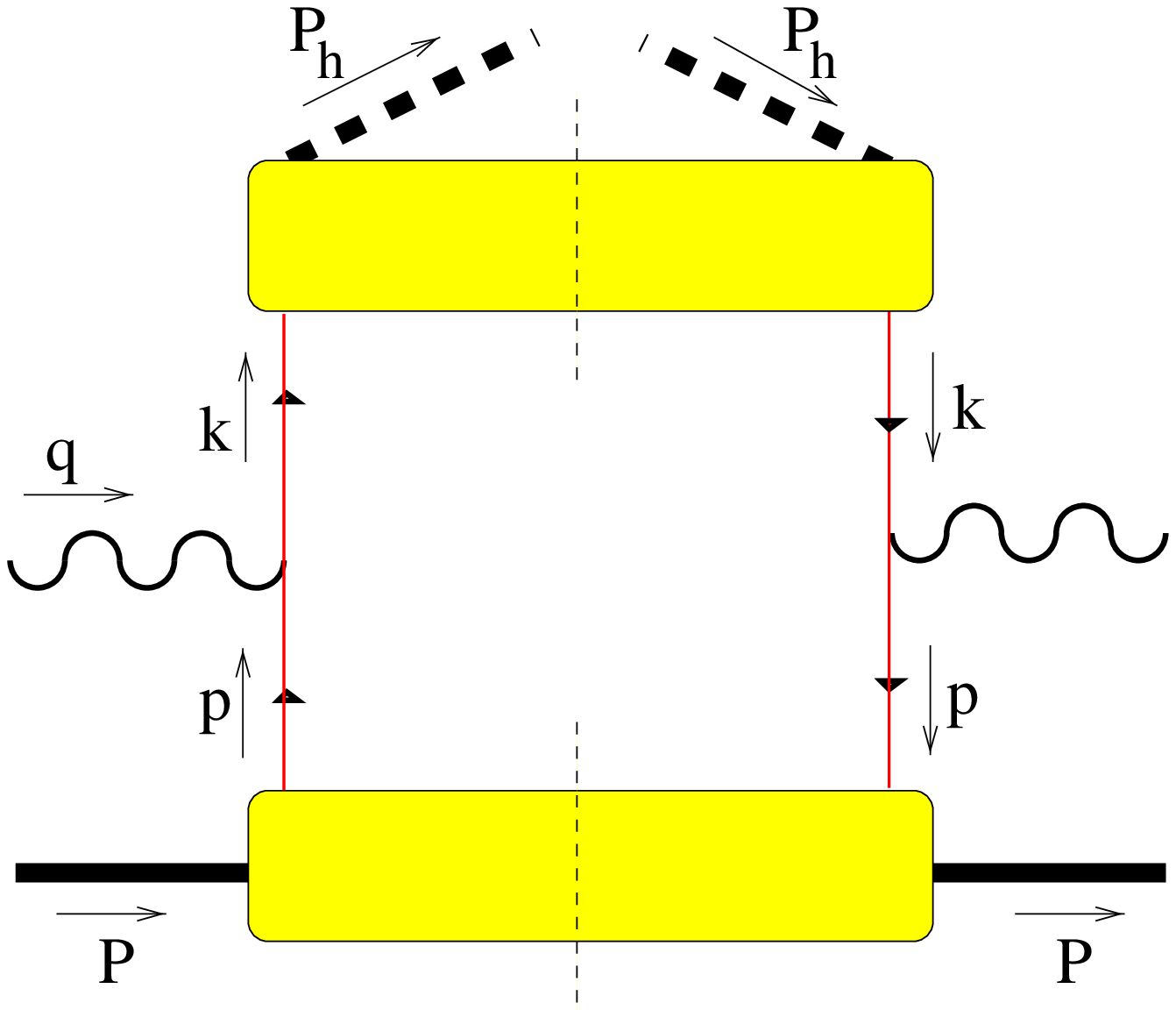,width=5cm} 
}
\caption{\label{figure1}
The leading order diagrams for inclusive lepton-hadron scattering
(left) and for semi-inclusive lepton-hadron scattering (right).}
\end{figure}
In the hard scattering approach the cross section for lepton-hadron 
scattering is for DIS in leading order given by
the left diagram in figure~\ref{figure1} representing the squared
amplitude (+ a similar antiquark contribution), while for SIDIS
the cross section is given by the right diagram in 
Fig.~\ref{figure1} (again + similar antiquark distribution). 
It is these contributions that will be analyzed in a number of cases.

\section{Quark distribution functions}

The first soft part to consider is the one that defines the quark
distribution functions. In a hard process such as leptoproduction
one can introduce two lightlike vectors, $n_+$ and $n_-$, satisfying 
$n_+^2 = n_-^2 =0$ and $n_+\cdot n_-$ = 1. The hadron momenta in the hard
scattering process can be taken proportional to one of the lightlike
vectors up to mass terms that are small compared to the hard scale 
in leptoproduction, the four momentum squared of the virtual photon, 
$q^2 = -Q^2$. We will assume $P \propto Q\,n_+$ 
and $P_h \propto Q\,n_-$. The lightlike vectors are used to define lightcone
coordinates $a^\pm \equiv a\cdot n_\mp$. The connection of hadron momenta
with the momenta of quarks and gluons is made via a soft part in which
all invariants, $p^2 \sim P\cdot p \sim P^2 = M^2 \ll Q^2$. This implies
that all $-$ components of momenta in the soft distribution part
are ${\cal O}(1/Q)$. The momenta
in the hard part have large $+$ {\em and} large $-$ components.

\begin{figure}
\centerline{
\epsfig{figure=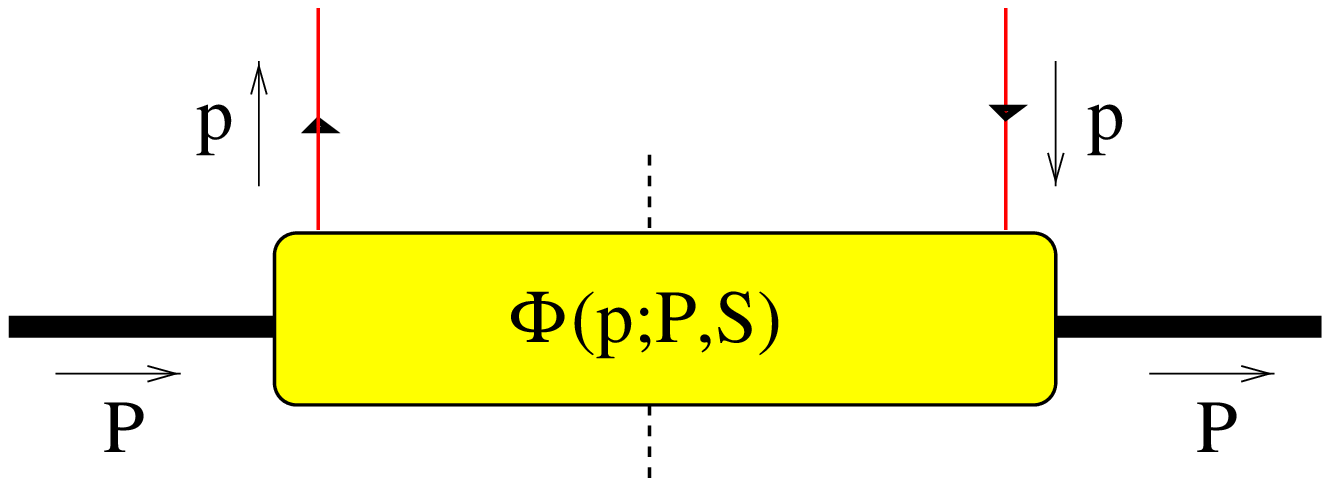,width=5.5cm}
\hspace{2cm}
\epsfig{figure=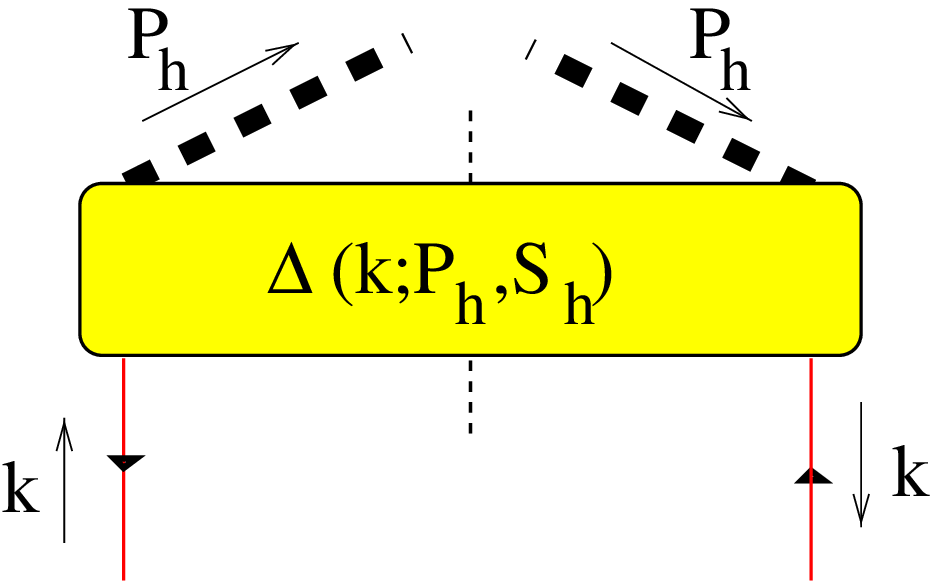,width=4cm}
}
\caption{\label{figure2}
The soft parts describing the {\em quark distribution} (left) and
{\em quark fragmentation} (right).}
\end{figure}
In DIS the only relevant component of
the quark momentum in the soft distribution part (Fig.~\ref{figure2})
is then the component $p^+ \equiv x\,P^+$.
Integrating over the other components of the quark momentum, the soft
part is (in the lightcone gauge $A^+ = 0$) given by
\be
\Phi_{ij}(x) = \left. \int \frac{d\xi^-}{2\pi}
\ e^{ip\cdot \xi} \,\langle P,S \vert \overline \psi_j (0)
\psi_i(\xi) \vert P,S \rangle \right|_{\xi^+ = \xi_\st= 0}.
\label{intdis}
\ee
Using Lorentz invariance, hermiticity, parity (P) and time-reversal (T)
one finds that in leading order in $1/Q$ it can be expanded as
\be
\Phi(x) =
\frac{1}{2}\,\left\{
f_1(x)\,\slash \slash n_+
+\lambda\,g_{1}(x)\, \gamma_5\slash n_+
+h_{1}(x)\,\gamma_5\,\frac{[\slash S_\st,n_+ ]}{2}
\right\} + {\cal O}\left(\frac{M}{P^+}\right),
\ee
where $\lambda = MS^+/P^+$ is the (lightcone) helicity and $S_\st$ is the 
transverse spin vector of the (spin 1/2) target 
(satisfying $\lambda^2 + \bm S_\st^2$ = 1). 
The quantities $f_1$, $g_1$ and $h_1$ can be readily interpreted
as densities (or differences thereof) for unpolarized quarks, longitudinally
polarized or transversely polarized quarks in a polarized target. Using
$\Phi$ to calculate the cross sections for lepton-hadron scattering one
obtains the familiar cross sections,
\bea
&&\frac{d\sigma_{OO}}{d\xbj dy}
= \frac{2\pi \alpha^2\,s}{Q^4}\,\xbj
\left\lgroup 1 + (1-y)^2\right\rgroup
\underbrace{\sum_{a,\bar a} e_a^2 \,f^a_1(\xbj)\ \mbox{}}_{
2\,F_1(\xbj) = F_2(\xbj)/\xbj},
\\ && \frac{d\sigma_{LL}}{d\xbj dy}
= \frac{2\pi \alpha^2\,s}{Q^4}\,\xbj y (2-y)\,\lambda_e\,\lambda
\,\underbrace{\sum_{a,\bar a} e_a^2 \,g^a_1(\xbj)\ \mbox{}}_{
2\,g_1(\xbj)},
\eea
where $\xbj = Q^2/2P\cdot q$ and $y = P\cdot q/P\cdot k$ are the usual
invariants. From the result one reads off, as indicated, the expressions 
for the structure functions $F_1$, $F_2$ in the case of unpolarized 
leptons scattering off an unpolarized target ($OO$). They are given by
the quark distributions $f_1^a$ weighted with the quark charges squared
and summed over quarks and antiquarks. Similarly one reads of the result
for $g_1$ in the case of longitudinally polarized leptons scattering
off a longitudinally polarized spin 1/2 target ($LL$).

\section{Fragmentation functions}

Next we turn to SIDIS. In the simplest case in which
one detects one hadron belonging to the current jet and determines in
essence only its longitudinal momentum, i.e. measures in addition to
$\xbj$ and $y$ the variable $z_h$ = $P\cdot P_h/P\cdot q$, one needs
only to consider the dependence on the component $k^- \equiv P_h^-/z$ in
the soft fragmentation part (Fig.~\ref{figure2}).
Integrating over the other components of $k$ the soft part is then
(in lightcone gauge $A^-$ = 0) given by
\be
\Delta(z) =
\left. \sum_X z\int \frac{d\xi^+}{4\pi} \
e^{ik\cdot \xi} \,\langle 0 \vert \psi (\xi) \vert X;P_h,S_h\rangle
\langle X;P_h,S_h\vert \overline \psi(0) \vert 0 \rangle
\right|_{\xi^- = \xi_\st = 0}.
\ee
The part relevant in $\Delta$ at leading order in $1/Q$ is
\be
\Delta(z) = z \left\{
D_1(z)\,\slash n_-
+ \lambda_h\,G_{1}(z)\,\gamma_5 \slash n_-
+ H_{1}(z)\,\gamma_5\,\frac{[\slash S_{h\st},\slash n_- ]}{2}
\right\} + {\cal O}\left(\frac{M_h}{P_h^-}\right),
\ee
where $\lambda_h = M_h\,S_h^-/P_h^-$ and $S_{h\st}$ determine the
polarization of the detected hadron. The fragmentation functions
$D_1$, $G_1$ and $H_1$ can be directly interpreted as quark
decay functions describing the decay for unpolarized, longitudinally
polarized or transversely polarized quarks into a polarized hadron.
Using $\Phi$ and $\Delta$ in Fig.~\ref{figure1} to calculate the cross 
section for SIDIS one obtains
\bea
&&\frac{d\sigma_{OO}}{d\xbj dy\,dz_h}
= \frac{2\pi \alpha^2\,s}{Q^4}\,\xbj
\left\lgroup 1 + (1-y)^2\right\rgroup
\,\sum_{a,\bar a} e_a^2 \,f^a_1(\xbj) \, D^a_1(z_h),
\label{unpol}
\\ && \frac{d\sigma_{LL}}{d\xbj dy\,dz_h}
= \frac{2\pi \alpha^2\,s}{Q^4}\,\xbj y (2-y)\,\lambda_e\,\lambda
\,\sum_{a,\bar a} e_a^2 \,g^a_1(\xbj)\,D^a_1(z_h),
\label{pol}
\eea
where the structure functions are given by products of distribution 
and fragmentation functions.

\begin{figure}
\centerline{
\epsfig{figure=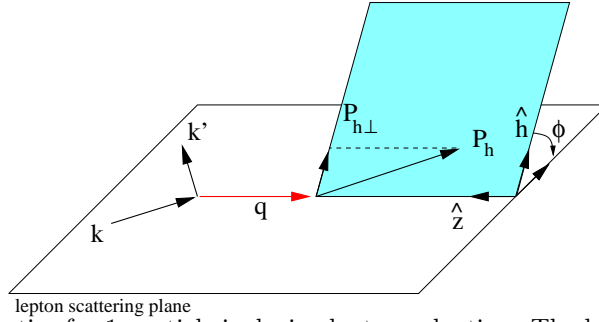,width=8cm}
}
\caption{\label{figure3}
Kinematics for 1-particle inclusive leptoproduction. The lepton scattering
plane is determined by lepton momenta and the hadron momentum $P$.}
\end{figure}
We note that the cross section for SIDIS in principle
can depend in additional to the variables $\xbj$, $y$ and $z_h$ on the
transverse momentum of the produced hadron, denoted $P_{h\perp}$ in the
frame where $P$ and $q$ do not have transverse components. Theoretically
it is convenient to work with
\be
q_\st^\mu 
\ = \ q^\mu + \xbj\,P^\mu - \frac{P_h^\mu}{\zh} \ =\ -\frac{P_{h\perp}}{\zh}
\ \equiv \ -Q_\st\,\hat h^\mu,
\ee
which is the transverse component of $q$ in a frame in which $P$ and
$P_h$ do not have transverse momenta. This vector is orthogonal to $n_+$ and
$n_-$. We will consider in the remainder cross sections 
dependending on $\xbj$, $y$ and $z_h$, but obtained after
weighting the full cross section with some function that may depend
on azimuthal angles as defined in Fig.~\ref{figure3},
\be
\langle W\rangle_{ABC}
= \int d\phi^\ell\,d^2\bm q_\st
\ W\,\frac{d\sigma_{ABC}^{[\vec e\vec H\rightarrow e\vec hX]}}
{d\xbj\,dy\,dz_h\,d\phi^\ell\,d^2\bm q_\st},
\label{asym}
\ee
where $W$ = $W(Q_\st,\phi_h^\ell,\phi_S^\ell,\phi_{S_h}^\ell)$.
In order to see in a glance which polarizations
are involved, we have added the subscripts ABC for polarizations of
lepton, target hadron and produced hadron, respectively. The cross
section in Eq.~\ref{unpol} is then denoted $\langle 1\rangle_{OOO}$,
that in Eq.~\ref{pol} as $\langle 1\rangle_{LLO}$.

\section{Polarimetry in SIDIS}

As an example of a weighted cross section
consider the process $\ell + H^\uparrow \rightarrow
\ell + h^\uparrow + X$ (e.g. $ep^\uparrow \rightarrow e\Lambda^\uparrow X$)
in which a spin 1/2 target is transversely polarized and one looks for 
transversely polarized spin 1/2 hadrons in the final state. The cross 
section using the expressions for $\Phi$ and $\Delta$ in the diagram in
Fig.~\ref{figure1} gives
\bea
&&
\left< \cos(\phi^\ell_S+\phi^\ell_{S_h})\right>_{OTT}
\nonumber \\ && \qquad \mbox{}
= \frac{2\pi \alpha^2\,s}{Q^4}
\,\vert \bm S_\st\vert\,\vert \bm S_{h\st} \vert
\left(1-y\right)\, \sum_{a,\bar a} e_a^2
\,\xbj\,h_1^a(\xbj) H^a_1(z_h).
\eea
This weighted cross section~\cite{Artru91} is the nonvanishing transverse
spin correlation between target hadron and produced hadron, probed via
the scattering off a transversely polarized quark, schematically
$target^\uparrow \Longrightarrow quark^\uparrow \Longrightarrow 
hadron^\uparrow$. Both the (transversely polarized) quark distribution
and the quark fragmentation function are chirally odd~\cite{JJ92}.

\section{Transverse momentum dependent quark distributions}

In the previous example, the transverse momentum of the outgoing hadron,
did not play a role. If measured, it requires consideration of transverse 
momenta in the soft part. Instead of Eq.~\ref{intdis} one needs
\be
\Phi(x,\bm p_\st) = \left. \int \frac{d\xi^-d^2\bm \xi_\st}{(2\pi)^3}
\ e^{ip\cdot \xi} \,\langle P,S \vert \overline \psi (0)
\psi(\xi) \vert P,S \rangle \right|_{\xi^+ = 0},
\ee
for which the relevant part in leading order is given 
by~\cite{RS79,TM95}
\bea
\Phi(x,\bm p_\st) &=&
\frac{1}{2}\,\Biggl\{
f_1\,\slash \slash n_+
+g_{1s}\, \gamma_5\slash n_+
\nonumber \\ & & \qquad
+h_{1T}\,\gamma_5\,\frac{[\slash S_\st , \slash n_+ ]}{2}
+h_{1s}^\perp\,\gamma_5\,\frac{[\slash p_\st,\slash n_+]}{2M}
\Biggr\},
\label{phieven}
\eea
with arguments $f_1$ = $f_1(x,\bm p_\st^2)$ etc. The quantity $g_{1s}$
(and similarly $h_{1s}^\perp$) is shorthand for
\be
g_{1s}(x,\bm p_\st) =
\lambda\,g_{1L}(x,\bm p_\st^2) + \frac{\bm p_\st\cdot \bm S_\st}{M}
\,g_{1T}(x,\bm p_\st^2).
\ee
The two functions $g_{1L}$ and $g_{1T}$ are interpreted as the quark
helicity distribution in a longitudinally and transversely polarized
target, respectively. Integrating over $\bm p_\st$ only $g_1(x)$ =
$\int d^2p_\st\ g_{1L}(x,\bm p_\st^2)$ survives. In the $p_\st$-weighted
result, $\Phi_\partial^\alpha$ = $\int d^2p_\st\ p_\st^\alpha\,\Phi$
only the $\bm p_\st^2$-moment defined as $g_{1T}^{(1)}(x)$ = 
$\int d^2p_\st\ (\bm p_\st^2/2M^2)\,g_{1T}(x,\bm p_\st^2)$ survives. 
This function appears for instance in the following weighted cross
section~\cite{K95,TM95a} in $\vec \ell + H^\uparrow \rightarrow 
\ell + h + X$ (e.g.\ $\vec ep^\uparrow \rightarrow e\pi^+X$),
\bea
&&
\left<\frac{Q_T}{M} \,\cos(\phi^\ell_h-\phi^\ell_S)\right>_{LTO}
\nonumber \\ && \qquad \mbox{}
= \frac{4\pi \alpha^2\,s}{Q^4}\,\lambda_e\,\vert \bm S_\st \vert
\,y\left(1-\frac{y}{2}\right) \sum_{a,\bar a} e_a^2
\,\xbj\,g_{1T}^{(1)a}(\xbj) D^a_1(z_h).
\eea
This weighted cross section correlates the transverse polarization
of the target with the azimuthal distribution of unpolarized hadrons
via the scattering of a longitudinally polarized quark, 
$target^\uparrow \Longrightarrow quark^\rightarrow \Longrightarrow
unpolarized$ $hadron$.

\section{Extension to subleading order}

When going to subleading order, i.e.\ $1/Q$ contributions in the cross
section, it is necessary to use for $\Phi$ the parametrization up to
$1/P^+$,
\bea
\Phi(x) & = &
\frac{1}{2}\,\Biggl\{f_1(x)\,\slash n_+
+ \lambda\,g_{1}(x)\,\gamma_5\slash n_+
+ h_{1}(x)\,\gamma_5\,\frac{[\slash S_\st,\slash n_+]}{2}
\Biggr\} \nonumber
\\ && \mbox{}+\frac{M}{2P^+}\Biggl\{
e(x) + g_T(x)\,\gamma_5\slash S_\st
+ \lambda\,h_L(x)\,\gamma_5\,\frac{[\slash n_+,\slash n_-]}{2}
\Biggr\}
\nonumber \\ && \mbox{} 
+ {\cal O}\left(\frac{M^2}{(P^+)^2}\right).
\eea
The twist-three functions $e$, $g_T$ and $h_L$ do not have a simple 
partonic interpretation. From the Lorentz structure of $\Phi$ and the
constraints imposed on it by Hermiticity, P and T one obtains 
(at tree-level)
relations~\cite{BKL84,MT96} with the transverse momentum dependent 
functions discussed in the previous section,
\bea
&&
\underbrace{g_T - g_1}_{g_2} \ =\  \frac{d}{dx}\,g_{1T}^{(1)}
\\ &&
\underbrace{h_L - h_1}_{{1\over 2}\,h_2}
\ =\  -\,\frac{d}{dx}\,h_{1L}^{\perp(1)}
\eea
The first relation has been used to get an estimate of $g_{1T}$ from the
$g_2$-data~\cite{KM96}.

\begin{figure}
\centerline{
\epsfig{figure=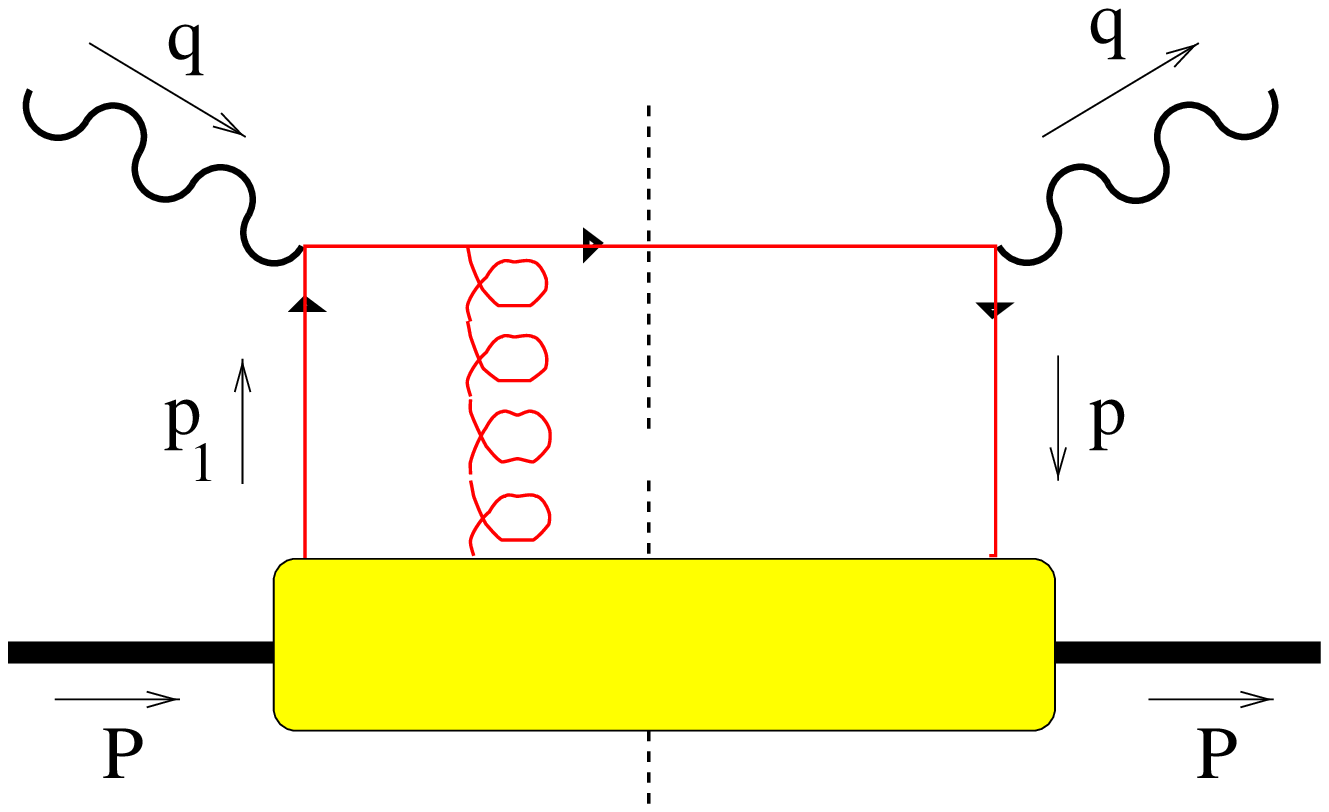,width=5cm}
\hspace{2cm}
\epsfig{figure=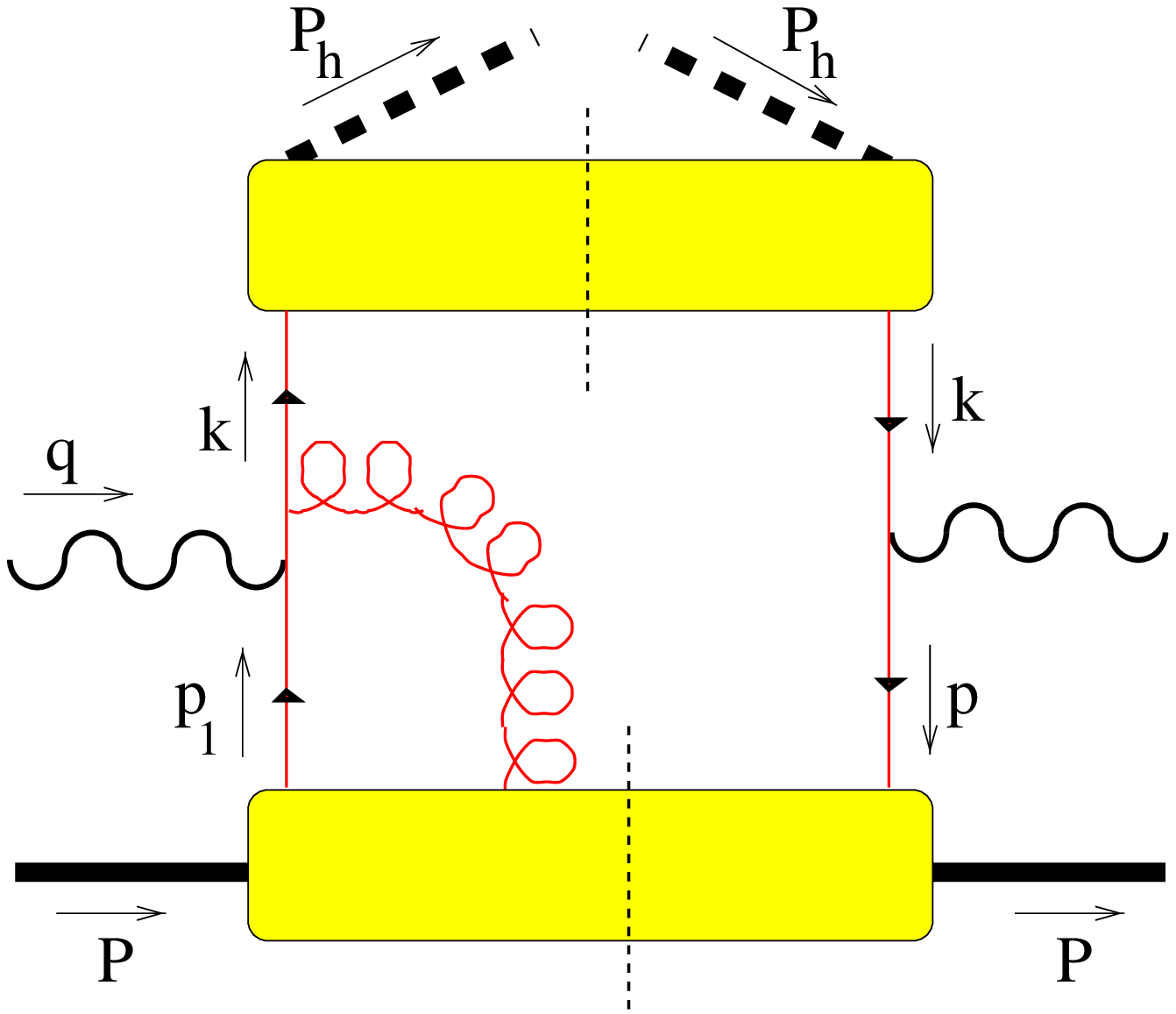,width=5cm} 
}
\caption{\label{figure4}
Examples of diagrams for DIS (left) and for SIDIS (right) needed at subleading
order.}
\end{figure}
At subleading order one needs to include soft parts containing gluon
fields~\cite{JJ92} as shown in Fig.~\ref{figure4}. 
These, however, can be dealt with via the QCD equations
of motion and they do not introduce new functions. Their contribution is
important to obtain a gauge invariant result. The most well-known example
is the structure function $g_2$ in DIS, or given as a
weighted cross section,
\be
\left<\cos(\phi^\ell_S)\right>_{LT}
= -\frac{4\pi \alpha^2\,s}{Q^4} \,\lambda_e\,\vert \bm S_\st\vert
\,y\sqrt{1-y} \,\sum_{a,\bar a} e_a^2
\,\frac{M\xbj^2}{Q}\,g_T^{a}(\xbj).
\ee

\section{T-odd fragmentation functions}

In the soft part discussed for the quark fragmentation, no constraints
arise from T invariance, because the states $\vert P_h, X\rangle$ are
out-states, which change into in-states under T. This allows additional
fragmentation functions in the case that transverse momentum is taken
into account. Restricting ourselves to {\em unpolarized} hadrons, we can
see that in the $\bm p_\st$-dependent distribution part only
one function $f_1(x,\bm p_\st^2)$ remains. The relevant part in $\Delta$ 
in leading order for unpolarized final states contains {\em two} functions,
\be
z\,\Delta(z,\bm k_\st) = D_1\,\slash n_-
+ H_1^\perp\,\frac{i\,[\slash k_\st, \slash n_-]}{2M_h}
+ {\cal O}\left(\frac{M_h}{P_h^-}\right),
\ee
with arguments $D_1$ = $D_1(z,z^2\bm k_\st^2)$ etc. Note that 
$\bm k_\st^\prime$ = $-z\,\bm k_\st$ is the transverse momentum of the
produced hadron with respect to the quark.

It turns out that the socalled T-odd functions (in this case $H_1^\perp$)
lead to single spin asymmetries~\cite{Co93,K95,TM95a,BM98}. 
For example the above function appears
in the production of unpolarized hadrons in leptoproduction in the case
of unpolarized leptons and a (longitudinally or transversely) polarized 
target, e.g.\ $ep^\uparrow \rightarrow e\pi^+X$,
\bea
&&
\left<\frac{Q_T}{M_h}\,\sin(\phi^\ell_h+\phi^\ell_S)\right>_{OTO}
\nonumber \\ && \qquad \mbox{}
= \frac{4\pi \alpha^2\,s}{Q^4}\,\vert \bm S_\st \vert
(1-y)\sum_{a,\bar a} e_a^2 \,\xbj h_1^a(\xbj)\,H_1^{\perp(1)a}(z_h),
\label{asym1}
\\&&
\left<\frac{Q_T^2}{4MM_h}\,\sin(2\phi^\ell_h)\right>_{OLO}
\nonumber \\ && \qquad \mbox{}
= -\frac{4\pi \alpha^2\,s}{Q^4}\,\lambda \,(1-y)\sum_{a,\bar a} e_a^2
\,\xbj h_{1L}^{\perp(1)a}(\xbj)\,H_1^{\perp(1)a}(z_h).
\eea
The interpretation is a correlation between the (transverse or 
longitudinal) polarization of the target and the azimuthal
distribution of the produced unpolarized hadrons, probed via scattering
off a transversely polarized quark, 
$target^{\uparrow /\rightarrow} \Longrightarrow quark^\uparrow$
$\stackrel{\mbox{\scriptsize T-odd}}{\Longrightarrow}$ $unpolarized$ 
$hadron$.
The same fragmentation function actually also appears in the scattering of
a polarized lepton from an unpolarized target~\cite{LM94}, but in that case 
appears in a subleading $\sin \phi_h^\ell$ asymmetry proportional
to $e(\xbj)\,H_1^{\perp (1)}(z_h)$.

\section{T-odd distribution functions}

For the distribution functions, it has been conjectured that T-odd
quantities also might appear
without violating time-reversal invariance~\cite{s90,abm95,alm96,adm96}.
This might be due to soft initial state interactions or, as suggested
recently~\cite{adm96}, be a consequence of chiral symmetry breaking.
Within QCD a possible description of the effects may come from gluonic
poles~\cite{BMT98}.
Here, let's simply assume the functions exist~\cite{BM98} in which case
Eq.~\ref{phieven} is extended with
\be
\Phi(x,\bm p_\st) = \ldots +
\frac{1}{2}\,\Biggl\{
f_{1T}^\perp\, \frac{\epsilon_{\mu \nu \rho \sigma}
\gamma^\mu n_+^\nu p_\st^\rho S_\st^\sigma}{M}
+ h_1^\perp\,\frac{i\,[\slash p_\st,\slash n_+]}{2M}
\Biggr\},
\label{phiodd}
\ee
A single spin asymmetry in which the function $f_{1T}^\perp$ appears 
is in the process
$\ell + H^\uparrow \rightarrow \ell + h + X$ (e.g.\ $ep^\uparrow
\rightarrow e\pi^+X$). One finds
\bea
&&
\left<\frac{Q_T}{M_h}\,\sin(\phi^\ell_h-\phi^\ell_S)\right>_{OTO}
\nonumber \\ &&\qquad \mbox{}
= \frac{2\pi \alpha^2\,s}{Q^4}\,\vert \bm S_\st \vert
\,\left( 1-y-\frac{1}{2}\,y^2\right) \sum_{a,\bar a} e_a^2
\,\xbj\,f_{1T}^{\perp (1)a}(\xbj) D_1^{a}(z_h).
\eea
This asymmetry is interpreted as a correlation between the transverse
polarization of the target and the azimuthal distribution of produced
hadrons via scattering off an unpolarized quark, $target^\uparrow$
$\stackrel{\mbox{\scriptsize T-odd}}{\Longrightarrow}$ 
$unpolarized\ quark$ $\Longrightarrow$
$unpolarized\ hadron$. Note that this is not the only single spin 
asymmetry for the OTO case (see Eq.~\ref{asym1}).

Finally, we note the interesting possibility that a combination of
T-odd distribution and fragmentation functions appears in unpolarized
scattering. This is the case for the $\cos(2\phi_h^\ell)$ asymmetry
in leptoproduction,
\bea
&&\left<\frac{Q_T^2}{4MM_h} \,\cos(2\phi^\ell_h)\right>_{OOO}
\nonumber \\ &&\qquad \mbox{}
= \frac{4\pi \alpha^2\,s}{Q^4} \,(1-y)
\sum_{a,\bar a} e_a^2
\,\xbj\,h_1^{\perp(1)a}(\xbj)\,H_1^{\perp (1)a}(z_h),
\eea
interpreted as $unpolarized\ target$
$\stackrel{\mbox{\scriptsize T-odd}}{\Longrightarrow}$
$quark^\uparrow$
$\stackrel{\mbox{\scriptsize T-odd}}{\Longrightarrow}$
$unpolarized\ hadron$.
This is a leading asymmetry, which in the absence of T-odd distribution
functions would start at order $1/Q^2$ (twist 4). 

\section{Summary}

In this talk many new possibilities have been outlined to probe the quark
and gluon structure of hadrons. The emphasis was on the transverse momentum
dependence in distribution and fragmentation functions that appear in 
semi-inclusive deep inelastic scattering at leading order. Some of these
functions or to be more precise their $\bm p_\st^2$-moments 
(e.g. $g_{1T}^{(1)}$ and $h_{1L}^{\perp (1)}$) are at tree-level simply 
related to twist-three functions (such as $g_T$ and $h_L$). Finally, 
the systematic
investigation of the soft parts formalizes many effects, such as the
Collins effect or final state interactions in leptoproduction.

\end{document}